\documentclass[12pt]{article}

\usepackage[margin=1in,footskip=0.25in]{geometry}

\usepackage{amsmath,amsfonts}
\usepackage{bm}
\usepackage{graphicx}
\usepackage{amssymb}
\usepackage{algorithm}
\usepackage{algpseudocode}
\usepackage{tikz}
\usepackage[caption=false]{subfig}
\usepackage{longtable,booktabs}
\usepackage{natbib}
\usepackage{hyperref}

\hypersetup{breaklinks=true,
            bookmarks=true,
            pdfauthor={J. Derek Tucker () and John R. Lewis () and Anuj Srivastava (Sandia National Laboratories and Florida State University)},
             pdfkeywords = {Compositional noise, functional data analysis, functional Principal
Component Analysis, functional regression},
            pdftitle={Elastic Functional Principal Component Regression},
            colorlinks=true,
            citecolor=blue,
            urlcolor=blue,
            linkcolor=magenta,
            pdfborder={0 0 0}}

\newcommand{\tex}[1]{#1}

\def\y{{\mathbf{y}}}

\def\F{{\mathcal{F}}}

\def\R{{\mathcal{R}}}

\DeclareMathOperator*{\argmax}{arg\,max}
\DeclareMathOperator*{\argmin}{arg\,min}

\newcommand{\s}{\ensuremath{\mathbb{S}}}
\newcommand{\real}{\mathbb{R}}
\newcommand{\fspace}{\ensuremath{\mathcal{F}}}
\newcommand{\T}{\ensuremath{\mathsf{T}}}

\newcommand{\ltwo}{\ensuremath{\mathbb{L}^2}}
\newcommand{\inner}[2]{\left\langle#1,#2 \right\rangle}
\numberwithin{equation}{section}

\usetikzlibrary{arrows,shapes,positioning,calc,fadings,decorations.pathreplacing,backgrounds}
\tikzset{%
  >=latex, 
  inner sep=0pt,%
  outer sep=2pt,%
  mark coordinate/.style={inner sep=0pt,outer sep=0pt, draw=blue!90,fill=blue!90,minimum size=4pt,circle},
  mark coordinate2/.style={inner sep=0pt,outer sep=0pt, draw=red!90,fill=red!90,minimum size=4pt,circle}%
}

\newcommand\pgfmathsinandcos[3]{%
  \pgfmathsetmacro{#1}{sin (#3)}%
  \pgfmathsetmacro{#2}{cos (#3)}%
}
\newcommand\LongitudePlane[3][current plane]{%
  \pgfmathsinandcos\sinEl\cosEl{#2} 
  \pgfmathsinandcos\sint\cost{#3} 
  \tikzset{#1/.style={cm={\cost,\sint*\sinEl,0,\cosEl, (0,0)}}}
}
\newcommand\LatitudePlane[3][current plane]{%
  \pgfmathsinandcos\sinEl\cosEl{#2} 
  \pgfmathsinandcos\sint\cost{#3} 
  \pgfmathsetmacro{\yshift}{\cosEl*\sint}
  \tikzset{#1/.style={cm={\cost,0,0,\cost*\sinEl, (0,\yshift)}}} %
}
\newcommand\DrawLongitudeCircle[2][1]{
  \LongitudePlane{\angEl}{#2}
  \tikzset{current plane/.prefix style={scale=#1}}
  \pgfmathsetmacro{\angVis}{atan (sin (#2)*cos (\angEl)/sin (\angEl))} %
  \draw[current plane,black!70] (\angVis:1) arc (\angVis:\angVis+180:1);
}
\newcommand\DrawLatitudeCircle[2][1]{
  \LatitudePlane{\angEl}{#2}
  \tikzset{current plane/.prefix style={scale=#1}}
  \pgfmathsetmacro{\sinVis}{sin (#2)/cos (#2)*sin (\angEl)/cos (\angEl)}
  \pgfmathsetmacro{\angVis}{asin (min (1,max (\sinVis,-1)))}
  \draw[current plane,black!80] (\angVis:1) arc (\angVis:-\angVis-180:1);
}

\begin{document}

\def\spacingset#1{\renewcommand{\baselinestretch}%
{#1}\small\normalsize} \spacingset{1}

\title{\bf Elastic Functional Principal Component Regression}

\author{J. Derek Tucker, John Lewis\thanks{This research was in part supported by the National Technical Nuclear
 Forensics Center (NTNFC) of the U.S. Department of Homeland Security
 (DHS). The authors would like to thank Dr.~Marc Welliver at SNL for his
 technical support during this work.}\hspace{.2cm}\\
Statistical Sciences, Sandia National Laboratories \\
and \\
Anuj Srivastava \\
Department of Statistics, Florida State University}
\maketitle

\begin{abstract}
  We study regression using functional predictors in situations where these functions contain both phase and amplitude variability. In other words, the functions are misaligned due to errors in time measurements, and these errors can significantly degrade both model estimation and prediction performance. The current techniques either ignore the phase variability, or handle it via pre-processing, i.e., use an off-the-shelf technique for functional alignment and phase removal. We develop a functional principal component regression model which has comprehensive approach in handling phase and amplitude variability. The model utilizes a mathematical representation of the data known as the square-root slope function. These functions preserve the $\ltwo$ norm under warping and are ideally suited for simultaneous estimation of regression and warping parameters. Using both simulated and real-world data sets, we demonstrate our approach and evaluate its prediction performance relative to current models. In addition, we propose an extension to functional logistic and multinomial logistic regression
\end{abstract}

\noindent%
\vfill

\newpage
\spacingset{1.45} 

\section{Introduction}\label{introduction}
The statistical analysis of functional data is fast gaining prominence
in the statistics community because ``big data" is central
to many applications. For instance, functional data can be found in
a broad swath of application areas ranging from biology, medicine, and
chemistry; to geology, sports, and financial analysis. In this problem,
some of the random quantities of interest are functions of independent
variables, (e.g., time, frequency), and are studied as elements of an
appropriate function space, often a Hilbert space. The analysis can
include common statistical procedures such as computing statistical
summaries, estimating parametric and nonparametric distributions, and
generating inferences under noisy observations. One common problem in
functional data analysis is regression modeling where the function
variables are used as predictors to estimate a scalar response variable.

More precisely, let the predictor functions be given by
\(\{f_i:[0,T] \rightarrow \real,~~i=1,2,\dots,n\}\) and the
corresponding response variables be \(y_i\). The standard functional
linear regression model for this set of observations is \begin{equation}
    y_i = \alpha + \int_0^T f_i(t)\beta(t)\,dt+\epsilon_i,~~i=1,\dots,n
    \label{eq:regress_model}
\end{equation} where \(\alpha \in \real\) is the intercept, \(\beta(t)\)
is the regression-coefficient function and \(\epsilon_i \in \real\) are
random errors. This model was first studied by
\citet{ramsay-dalzell:1991} and \citet{cardot-ferraty-sarda:1999}. The
model parameters are usually estimated by minimizing the sum of squared
errors (SSE),
\[\{\alpha^*,\beta^*(t)\} = \argmin_{\alpha,\beta(t)} \sum_{i=1}^n |y_i - \alpha - \int_0^T f_i(t)\beta(t)\,dt|^2.\]
These values form maximum-likelihood estimators of parameters under the
additive white-Gaussian noise model. One problem with this approach is
the solution is an element of an infinite-dimensional space while its
specification for any $n$ is finite dimensional; resulting in an infinite
number of possible solutions without further restrictions imposed on the problem.
\citet{ramsay-silverman-2005} proposed two
approaches to handle this issue: (1) Represent \(\beta(t)\) using \(p\)
basis functions in which \(p\) is kept large to allow desired variations
of \(\beta(t)\), and (2) add a roughness penalty term to the objective
function (SSE) which selects a smooth solution by finding an optimal
balance between the SSE and the roughness penalty. The basis can come
from Fourier analysis, splines, or functional PCA
(\citet{reiss-ogden:2007}).

Current literature in functional linear regression is focused primarily
on the estimation of the coefficient of \(\beta(t)\) under a basis
representation. For example,
\citep{cuevas-febrero-fraiman:2002, cardot-ferraty-sarda:2003, Hall2007, James2009}
discuss estimation and/or inference of \(\beta(t)\) for different cases
for the standard functional linear model and the interpretation of
\(\beta(t)\). Focusing on prediction of the scalar response,
\citet{Cai2006} studied the estimation of \(\int f_i(t)\beta(t)\,dt\).
In some situations the response variable \(y_i\), is categorical and the
standard linear model will not suffice. \citet{james:2002} extended the
standard functional linear model to functional logistic regression to be
able to handle such situations. \citet{muller-stadtmuller:2005} extended
the generalized model to contain a dimension reduction by using a
truncated Karhunen-Lo\tex{\`{e}}ve expansion. Recently,
\citet{gerheiss-maity-staicu:2013} included variable selection to reduce
the number of parameters in the generalized model.

In practice the predictor functions are observed at discrete points and
not the full interval \([0,T]\). Furthermore, in some situations, these
observations are corrupted by noise along the time axis. That is, one
observes \(\{(t + \eta(t), f(t))\}\) instead of \(\{(t, f(t))\}\) where
the random variables \(\eta(t)\) are constrained so that the observation
times do not cross each other. While some papers have assumed parametric
models for \(\eta(t)\) (\citet{carroll-etal:2006}) and incorporated them
in the estimation process, the others have ignored them completely. It
is more natural to treat these measurement variables in a nonparametric
form as follows: We assume that observation times are given by
\(\gamma(t)\) where \(\gamma\) is a monotonic function with appropriate
boundary conditions (\(\gamma(0) = 0,\ \gamma(T) = T\)). Consequently,
the observations are modeled as \(\{\gamma(t), f(t)\}\) where \(\gamma\)
captures a random noise component that needs to be accounted for in the
estimation process. The effect of \(\gamma\) is a warping of \(f\), with
a nonlinear shift in the \emph{locations} of peaks and valleys but no
changes in the heights of those peaks and valleys. In this effect warping differs
across realizations (observations) and, hence, is termed as
\emph{warping or compositional} noise. Some authors have also called it
the phase variability in functional data. If the phase variability is
ignored, the resulting model may fail to capture patterns present in the
data and will lead to inefficient data models. One way to handle this
noise is to capture both the phase and amplitude variability properly in
the regression model. It is more natural to include handling of warping
noise, or alignment, in the regression model estimation itself; and
perform a joint inference on all model variables under the same
objective function. Recently, \citet{gervini:2013} has proposed a
functional linear regression model that includes phase-variability in
the model. This uses a random-effect simultaneous linear model on the
warping parameters and the principal component scores (of aligned
predictor functions). However, this method involves PCA on the original
functional space and has shown to be inferior for unaligned data
(\citet{tucker-wu-srivastava:2013}:\citet{Lee:2017}).
\citet{tucker-wu-srivastava:2013} showed that if this variability is not
accounted for properly when performing fPCA, the results will be
misleading due to incorrect shape of the calculated mean function.

In this paper we focus on problems where the functional data contains
random phase variability. To handle that variability, we propose a
regression model that incorporates the phase variability through the use
of functional principal component regression (fPCR); where this
variability is handled in a harmonious way. The basic idea is to use a
PCA method as the basis that is able to capture the amplitude
variability, phase variability, or both in the regression problem. This
allows the model to capture the variability that is important in
predicting the outcome from the data. Using this representation and the
geometry of the warping function \(\gamma\), we construct the model and
outline the resulting prediction procedures. The fPCR method was first
proposed by \citet{reiss-ogden:2007}, but they neglected to account for the
phase-variability found in functional data. We extend this
framework to the logistic regression case where the response can take on
categorical data. We will illustrate this application using both
simulated and real data sets, which includes sonar, gait, and
electrocardiogram data. The physiological data is studied in the context
of classification of disease types or the separation of individuals.

This paper is organized as follows: In Section \ref{theory} we review
the relevant material from functional regression, and in Section
\ref{elastic} we develop the elastic functional PCR model. In Section
\ref{logistic} we extend the elastic fPCR to the logistic and
multinomial logistic case. In Sections \ref{sim} and \ref{realdata}, we
report the results of applying the proposed approach to a simulated data
set and seven real data sets from various application domains. Finally,
we close with a brief summary and some ideas for future work in Section
\ref{conclusion}.

\section{Functional Principal Component Regression}\label{theory}

We start with a more common functional regression model, and then
develop an ``elastic'' principal component version that accounts for
phase variability of the functional data. Without loss of generality we
assume the time interval of interest to be \([0,1]\). Let \(f\) be a
real-valued function on \([0,1]\); from a theoretical perspective we
restrict to functions that are absolutely continuous on \([0,1]\) and we
let \(\F\) denote the set of all such functions. In practice, since
observed data are discrete, this assumption is not a restriction.

\subsection{Functional Principal Component Regression Model}\label{functional-principal-component-regression-model}

Let \(\{f_i\}\) denote observations of a predictor function variable and
let \(y_i\in\real\), be the corresponding response variable. The
standard functional linear regression model for this set of observations
is \begin{equation}
    y_i = \alpha + \int f_i(t)\beta(t)\,dt+\epsilon_i,~~i=1,\dots,n
    \label{eq:regress_model}
\end{equation} where \(\alpha\) is the bias and \(\beta(t)\) is the
regression coefficient function. The model is usually determined by
minimizing the sum of squared errors (SSE). \begin{equation}
    \{\alpha^*,\beta^*(t)\} = \argmin_{\alpha,\beta(t)} \sum_{i=1}^n |y_i - \alpha - \int f_i(t)\beta(t)\,dt|^2.
    \label{eq:regress_SSE}
\end{equation} One problem with this approach, is that for any finite
\(n\), it is possible to perfectly interpolate the responses if no
restrictions were are placed on \(\beta(t)\). Specifically, since
\(\beta(t)\) is infinite dimensional, we have infinite degrees of
freedom to form \(\beta(t)\) in which we can make the SSE equal zero.
\citet{ramsay-silverman-2005} proposed two approaches with the first,
representing \(\beta(t)\) using a \(p\)-dimensional basis in which \(p\)
is hopefully large enough to capture all variations of \(\beta(t)\). The
second approach is adding a penalty term which shrinks the variability
of \(\beta(t)\) or smooths its response.

Functional principal component regression uses the principal components
as the basis functions where the model is determined by minimizing
\begin{equation}
    \{\alpha^*,\mathbf{b}^*\} = \argmin_{\alpha,\mathbf{b}} \sum_{i=1}^n |y_i - \alpha - \sum_{j=1}^{n_o} \inner{f_i(t)}{\xi_j(t)}b_j|^2,
    \label{eq:fPCR_SSE}
\end{equation} where \(n_o\) principal components are used, \(\xi(t)\)
is the corresponding eigenfunction, and
\(\mathbf{b} = [b_1,\dots,b_j]\). It should be noted that \(f_i(t)\)
here is mean centered.

\section{Elastic Functional Principal Component Regression
Model}\label{elastic}

In order to properly account for the variability, we can use the vertical
fPCA and horizontal fPCA presented in \citet{tucker-wu-srivastava:2013}.
These PCA methods account for the variability, by first separating the
phase and amplitude and then performing the PCA on the spaces
separately. Using these methods, one can construct a regression on the
amplitude space using the square-root slope function (SRSF), \(q\) and
specifically the aligned SRSF or the phase space using the warping
functions, \(\gamma\), the motivation for the using of SRSF will be
explained later. A third option is to use the method developed by
\citet{Lee:2017} which is an extension of the method developed by Tucker
et al.~Lee proposes a combined fPCA which generates a function \(g^C\),
which combines the function (\(f\)) and the warping function. We propose
a slight modification to this work to use the SRSF, due to its
theoretical properties. The combined function does work on a
simplified geometry of the warping function where the warping function,
is transformed to the Hilbert Sphere; and the shooting vector that maps
to the tangent space is analyzed. This simplification, and SRSF
modification allows the use of a metric that is a proper distance as in
the vertical and horizontal case. By using the combined fPCA the
regression model can be performed on the amplitude and phase
simultaneously. Table \ref{tab:domains} presents the three domains and
where the regression is performed.

\begin{table*}[!h]

\caption{\label{tab:domains}Functional Principal Component Regression Domains.}
\centering
\begin{tabular}[t]{lccc}
\toprule
  & Vertical fPCA & Horizontal fPCA & Combined fPCA\\
\midrule
Domain & $\tilde{q}$ & $\gamma$ & $g^C = [\tilde{q}~~Cv(t)]$\\
Variability & Amplitude & Phase & Amplitude + Phase\\
Metric & Fisher-Rao & Fisher-Rao & Fisher-Rao\\
\bottomrule
\end{tabular}
\end{table*}

\subsection{Elastic Functional fPCA}\label{pca}

We begin by giving a short review of the vertical and horizontal fPCA of
\citet{tucker-wu-srivastava:2013} and the combined phase-amplitude fPCA
method of \citet{Lee:2017}, with a slight modification which will be
described clearly in later sections. These methods are based on the
functional data analysis approach outlined in
\citet{srivastava-etal-JASA:2011},
\citet{kurtek-wu-srivastava-NIPS:2011}, and
\citet{tucker-wu-srivastava:2013}; see those references for more details
on this background material.

Let \(\Gamma\) be the set of orientation-preserving diffeomorphisms of
the unit interval \([0,1]\):
\(\Gamma = \{\gamma: [0,1] \rightarrow [0,1] |~\gamma(0) = 0,~\gamma(1)=1,\gamma~\text{is a diffeomorphism} \}\).
Elements of \(\Gamma\) play the role of warping functions. For any
\(f \in \F\) and \(\gamma \in \Gamma\), the composition
\(f \circ \gamma\) denotes the time-warping of \(f\) by \(\gamma\). With
the composition operation, the set \(\Gamma\) is a Lie group with the
identity element \(\gamma_{id}(t) = t\). This is an important
observation since the group structure of \(\Gamma\) is seldom utilized
in past papers on functional data analysis.

As described in \citet{tucker-wu-srivastava:2013}, there are two metrics
to measure the amplitude and phase variability of functions. These
metrics are proper distances, one on the quotient space \(\F/\Gamma\)
(i.e., amplitude) and the other on the group \(\Gamma\) (i.e., phase).
The amplitude or \(y\)-distance for any two functions
\(f_1,\ f_2 \in \F\) is defined as
\begin{equation}
d_a(f_1, f_2) = \inf_{\gamma \in \Gamma} \|q_1 - (q_2 \circ \gamma)\sqrt{\dot{\gamma}}\|,
\label{eq:d_a}
\end{equation} where
\(q(t) = \mbox{sign}(\dot{f}(t)) \sqrt{ |\dot{f}(t)|}\) is known as the
square-root slope function (SRSF) (\(\dot{f}\) represents the derivative
of \(f\)). The optimization problem in Equation \ref{eq:d_a} is most
commonly solved using a Dynamic Programming algorithm; see
\citet{robinson-2012} for a detailed description. If \(f\) is absolutely
continuous, then \(q\in\ltwo([0,1],\real)\) (\citet{robinson-2012}),
henceforth denoted by \(\ltwo\). For the properties of the SRSF and the
reason for its use in this setting, we refer the reader to
\citet{srivastava-kalseen-joshi-jermyn:11}, \citet{marron2015} and
\citet{LRK}. Moreover, it can be shown that for any
\(\gamma_1, \gamma_2 \in \Gamma\), we have
\(d_a(f_1 \circ \gamma_1, f_2 \circ \gamma_2) = d_a(f_1, f_2)\), i.e.,
the amplitude distance is invariant to function warping.

\subsection{\texorpdfstring{Simplifying Geometry of
\(\Gamma\)}{Simplifying Geometry of \textbackslash{}Gamma}}\label{simplifying-geometry-of-gamma}

The space of warping functions, \(\Gamma\), is an infinite-dimensional
nonlinear manifold, and therefore cannot be treated as a standard
Hilbert space. To overcome this problem, we will use tools from
differential geometry to perform statistical analyses and to model the
warping functions. The following framework was previously used in
various settings including; (1) modeling re-parameterizations of curves
(\citet{srivastava-jermyn-PAMI:09}), (2) putting prior distributions on
warping functions (\citet{Kurtek17} and \citet{Lu17}), (3) studying
execution rates of human activities in videos
(\citet{ashok-srivastava-etal-TIP:09}), and many others. It is also very
closely related to the square-root representation of probability density
functions introduced by \citet{bhattacharya-43}, and later used for
various statistical tasks (see e.g., \citet{SK}).

\begin{figure}[t]
\begin{center}
  \begin{tikzpicture}
  	\def\R{2.5} 
    \def\angEl{20} 
	\def\angAz{-55} 
	\def\angPhiOne{67} 
	\pgfmathsetmacro\H{\R*cos(\angEl)} 
	\tikzset{xyplane/.estyle={cm={cos(\angAz),sin(\angAz)*sin(\angEl),-sin(\angAz),
	                              cos(\angAz)*sin(\angEl),(0,\H)}}}
	\tikzset{dot/.style={circle,draw=red!90,fill=red!90,minimum size=3.5pt}}
	\tikzset{dot2/.style={circle,draw=blue!90,fill=blue!90,minimum size=3.5pt}}
	\LongitudePlane[xzplane]{\angEl}{\angAz}
	\LongitudePlane[pzplane]{\angEl}{\angPhiOne}
	\LatitudePlane[equator]{\angEl}{0}
	\node at (.15,\R-1) [dot] (J) {};
	\node at (.15,\R+.5) [dot2] (10) {};
	\draw[green!90,line width=1.2pt, shorten <=-3pt,shorten >=-3pt] (J) -- (10);
	\fill[ball color=gray!30] (0,0) circle (\R);
	\draw[xyplane,blue,line width=1.2pt] (-1.25*\R,-1*\R) rectangle (1.25*\R,1*\R);
	\coordinate (O) at (0,0);
	\coordinate (N) at (0,\H);
	\coordinate[label=above:$\psi_{id}$] (N1) at (-1.25,1.8+\H);
	\coordinate[label=above:$v_i$] (v) at (-3.25,.8+\H);
	\coordinate[label=above:$\psi_i$] (N2) at (-3.25,\H-1);
	\DrawLatitudeCircle[\R]{0};
	\DrawLatitudeCircle[\R]{15};
	\DrawLatitudeCircle[\R]{30};
	\DrawLatitudeCircle[\R]{45};
	\DrawLatitudeCircle[\R]{60};
	\DrawLatitudeCircle[\R]{75};
	\DrawLatitudeCircle[\R]{89};
	\DrawLatitudeCircle[\R]{-15};
	\DrawLatitudeCircle[\R]{-30};
	\DrawLatitudeCircle[\R]{-45};
	\DrawLongitudeCircle[\R]{160};
	\DrawLongitudeCircle[\R]{140};
	\DrawLongitudeCircle[\R]{120};
	\DrawLongitudeCircle[\R]{100};
	\DrawLongitudeCircle[\R]{80};
	\DrawLongitudeCircle[\R]{60};
	\DrawLongitudeCircle[\R]{40};
	\DrawLongitudeCircle[\R]{20};
	\node at (2,\R) [dot2] (1) {};
	\node at (.8,\R-.5) [dot2] (2) {};
	\node at (1.3,1.6) [dot2] (3) {};
	\node at (-1.4,\R+.59) [dot2] (4) {};
	\node at (-2.1,\R+.12) [dot2] (5) {};
	\node at (-1,\R-.22) [dot2] (6) {};
	\node at (.9,\R+.15) [dot2] (7) {};
	\node at (.3,\R+.1) [dot2] (8) {};
	\node at (.2,1.8) [dot2] (9) {};
	\node at (.3,\R-.05) [dot] (A) {};
	\node at (.9,.55) [dot] (B) {};
	\node at (1.8,1.6) [dot] (C) {};
	\node at (-1.3,2.1) (D) {};
	\node at (-2,1) [dot] (E) {};
	\node at (.8,1.7) [dot] (F) {};
	\node at (-1,2) [dot] (G) {};
	\node at (.85,\R-.25) [dot] (H) {};
	\node at (.2,1.3) [dot] (I) {};
	\draw[-latex,blue!70,thick] (N1) -- (N);
	\draw[-latex,blue!70,thick] (v) -- (5);
	\draw[-latex,blue!70,thick] (N2) -- (E);
	\draw[green!90,line width=1.2pt, shorten <=-3pt,shorten >=-3pt] (B) .. controls (1.5,1.4) .. (3);
	\draw[green!90,line width=1.2pt, shorten <=-3pt,shorten >=-3pt] (C) .. controls (\R-.3,2.3) .. (1);
	\draw[green!90,line width=1.2pt, shorten <=-3pt,shorten >=-3pt] (D) .. controls (-1.5,\R+.5) .. (4);
	\draw[green!90,line width=1.2pt, shorten <=-3pt,shorten >=-3pt] (E) .. controls (-2.3,\R-.1) .. (5);
	\draw[green!90,line width=1.2pt] (A) -- (8);
	\draw[green!90,line width=1.2pt, shorten <=-3pt,shorten >=-3pt] (F) -- (2);
	\draw[green!90,line width=1.2pt, shorten <=-3pt,shorten >=-3pt] (G) -- (6);
	\draw[green!90,line width=1.2pt, shorten <=-3pt,shorten >=-3pt] (H) -- (7);
	\draw[green!90,line width=1.2pt, shorten <=-3pt,shorten >=-3pt] (I) -- (9);
\end{tikzpicture}
  \caption{Depiction of the SRSF space of warping functions as a sphere and a tangent space at the identity element $\psi_{id}$.}
  \label{fig:sphere-map}
\end{center}
\end{figure}
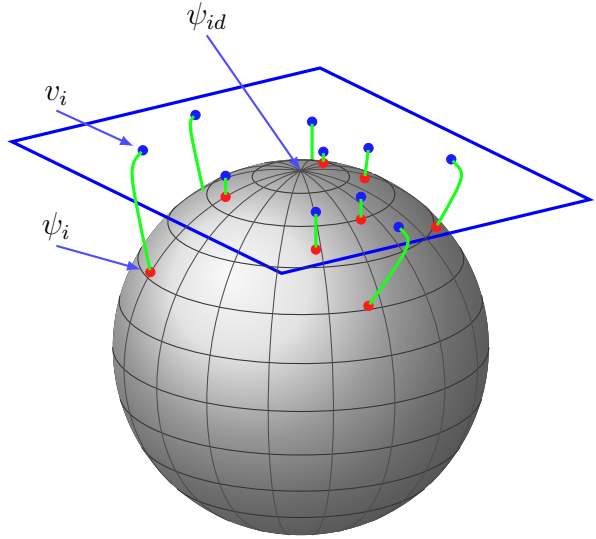

We represent an element \(\gamma \in \Gamma\) by the square-root of its
derivative \(\psi = \sqrt{\dot{\gamma}}\). Note that this is the same as
the SRSF defined earlier, and takes this form since
\(\dot{\gamma} > 0\). The identity \(\gamma_{id}\) maps to a constant
function with value \(\psi_{id}(t) = 1\). Since \(\gamma(0) = 0\), the
mapping from \(\gamma\) to \(\psi\) is a bijection, and one can
reconstruct \(\gamma\) from \(\psi\) using
\(\gamma(t) = \int_0^t \psi(s)^2 ds\). An important advantage of this
transformation is that since
\(\| \psi\|^2 = \int_0^1 \psi(t)^2 dt = \int_0^1 \dot{\gamma}(t) dt = \gamma(1) - \gamma(0) = 1\),
the set of all such \(\psi\)s is the positive orthant of the Hilbert
sphere \(\Psi=\s_{\infty}^+\) (i.e., a unit sphere in the Hilbert space
\(\ltwo\)). In other words, the square-root representation simplifies
the complicated geometry of \(\Gamma\) to a unit sphere. The distance
between any two warping functions, i.e., the phase distance, is exactly
the arc-length between their corresponding SRSFs on the unit sphere
\(\s_{\infty}\): \[
d_{p}(\gamma_1, \gamma_2) = d_{\psi}(\psi_1, \psi_2) \equiv \cos^{-1}\left(\int_0^1 \psi_1(t) \psi_2(t) dt \right)\ .
\] Figure \ref{fig:sphere-map} shows an illustration of the SRSF space
of warping functions as a unit sphere.

\subsection{Mapping to the Tangent Space at Identity
Element}\label{mapping-to-the-tangent-space-at-identity-element}

While the geometry of \(\Psi\subset\s_{\infty}\) is more tractable, it
is still a nonlinear manifold and computing standard statistics remains
difficult. Instead, we use a tangent (vector) space at a certain fixed
point for further analysis. The tangent space at any point
\(\psi \in \Psi\) is given by:
\(T_{\psi}(\Psi) = \{v \in \ltwo| \int_0^1 v(t) \psi(t) dt = 0\}\). To
map between the representation space \(\Psi\) and tangent spaces, one
requires the exponential and inverse-exponential mappings. The
exponential map at a point \(\psi\in\Psi\) denoted by
\(\exp_\psi : T_{\psi}(\Psi) \mapsto \Psi\), is defined as

\begin{equation}
\exp_\psi(v) = \cos(\|v\|)\psi+\sin(\|v\|)\frac{v}{\|v\|},
\end{equation} \noindent where \(v\in T_{\psi}(\Psi)\). Thus,
\(\exp_\psi(v)\) maps points from the tangent space at \(\psi\) to the
representation space \(\Psi\). Similarly, the inverse-exponential map,
denoted by \(\exp_{\psi}^{-1} : \Psi \mapsto T_{\psi}(\Psi)\), is
defined as

\begin{equation}
\exp_{\psi}^{-1}(\psi_1) = \frac{\theta}{\sin(\theta)}(\psi_1-\cos(\theta)\psi),
\end{equation} \noindent where \(\theta = d_{p}(\gamma_1, \gamma)\).
This mapping takes points from the representation space to the tangent
space at \(\psi\).

The tangent space representation \(v\) is sometimes referred to as a
\emph{shooting vector}, as depicted in Figure~\ref{fig:sphere-map}. The
remaining question is which tangent space should be used to represent
the warping functions. A sensible point on \(\Psi\) to define the
tangent space is the sample Karcher mean \(\hat{\mu}_{\psi}\)
(corresponding to \(\hat{\mu}_{\gamma}\)) of the given warping functions
or the identity element \(\psi_{id}\). For details on the definition of
the sample Karcher mean and how to compute it, please refer to
\citet{tucker-wu-srivastava:2013}.

\subsection{Vertical Functional Principal
Components}\label{vertical-functional-principal-components}

Let \(f_1,\cdots,f_n\) be a given set of functions, and
\(q_1,\cdots,q_n\) be the corresponding SRSFs, \({\mu}_q\) be their
Karcher Mean, and let \(\tilde{q}_i\)s be the corresponding aligned
SRSFs using Algorithm 1 from \citet{tucker-wu-srivastava:2013}. In
performing vertical fPCA, one needs to include the variability
associated with the initial values, i.e., \(\{f_i(0)\}\), of the given
functions. Since representing functions by their SRSFs ignores this
initial value, this variable is treated separately. That is, a
functional variable \(f\) is analyzed using the pair \((q, f(0))\)
rather than just \(q\). This way, the mapping from the function space
\(\fspace\) to \(\ltwo \times \real\) is a bijection. In practice, where
\(q\) is represented using a finite partition of \([0,1]\), say with
cardinality \(T\), the combined vector \(h_i = [q_i~~f_i(0)]\) simply
has dimension \((T+1)\) for fPCA. We can define a sample covariance
operator for the aligned combined vector
\(\tilde{h} = [\tilde{q}_1~~f_i(0)]\) as
\[K_h = \frac{1}{n-1}\sum_{i=1}^n E[(\tilde{h}_i - \mu_h)(\tilde{h}_i - \mu_h)^\T] \in \mathbb{R}^{(T+1) \times (T+1)}\ ,\]
where \(\mu_h = [\mu_q~~ \bar{f}(0)]\). Taking the SVD,
\(K_h=U_h\Sigma_h V_h^\T\) we can calculate the directions of principle
variability in the given SRSFs using the first \(p\leq n\) columns of
\(U_h\) and can be converted back to the function space \(\mathcal{F}\),
via integration, for finding the principal components of the original
functional data. Moreover, we can calculate the observed principal
coefficients as \(\inner{\tilde{h}_i}{U_{h,j}}\).

One can then use this framework to visualize the vertical
principal-geodesic paths. The basic idea is to compute a few points
along geodesic path
\(\tau \mapsto \mu_h + \tau\sqrt{\Sigma_{h,jj}}U_{h,j}\) for
\(\tau \in \real\) in \(\ltwo\), where \(\Sigma_{h,jj}\) and \(U_{h,j}\)
are the \(j^{th}\) singular value and column, respectively. Then, obtain
principle paths in the function space \(\fspace\) by integration.

\subsection{Horizontal Functional Principal
Components}\label{horizontal-functional-principal-components}

To perform horizontal fPCA we will use the tangent space at
\(\mu_{\psi}\) to perform analysis, where \(\mu_{\psi}\) is the mean of
the transformed warping functions. Algorithm 2 from
\citet{tucker-wu-srivastava:2013} can be used to calculate this mean. In
this tangent space we can define a sample covariance function:
\[K_\psi = \frac{1}{n-1} \sum_{i=1}^n E[v_i v_i^\T] \in \real^{T\times T}.\]
The singular value decomposition (SVD) of
\(K_{\psi} = U_{\psi} \Sigma_{\psi} V_{\psi}^\T\) provides the estimated
principal components of \(\{ \psi_i\}\): the principal directions
\(U_{\psi,j}\) and the observed principal coefficients
\(\inner{v_i}{U_{\psi,j}}\). This analysis on \(\s_{\infty}\) is similar
to the ideas presented in \cite{srivastava-joshi-etal:05} although one
can also use the idea of principal nested sphere for this analysis
\cite{jung-dryden-marron:2012}. The columns of \(U_{\psi}\) can then be
used to visualize the principal geodesic paths.

\subsection{Combined Functional Principal
Components}\label{combined-functional-principal-components}

To model the association between the amplitude of a function and its
phase, \citet{Lee:2017} use a combined function \(g^C\) on the extended
domain \([0,2]\) (for some \(C>0\))

\begin{equation}
g^C(t) = \left\{\begin{array}{l r} f^*(t), & t\in[0,1) \\ C v(t-1), & t\in[1,2]\end{array}\right.
\end{equation} \noindent where \(f^*\) only contains the function's
amplitude (i.e., after alignment via SRSFs). Furthermore,
\citet{Lee:2017} assume that \(g^C\in \ltwo([0,2],\real)\). The
parameter \(C\) is introduced to adjust for the scaling imbalance
between \(f^*\) and \(v\). In their current work, we make a slight
modification to the method of \citet{Lee:2017}. In particular, it seems
more appropriate to construct the function \(g^C\) using the SRSF
\(q^*\) of the aligned function \(f^*\), since \(q^*\) is guaranteed to
be an element of \(\ltwo\). Thus, with a slight abuse in notation, we
proceed with the following joint representation of amplitude and phase:
\begin{equation}\label{eq:combfun}
g^C(t) = \left\{\begin{array}{l r} q^*(t), & t\in[0,1) \\ C v(t-1), & t\in[1,2]\end{array}\right.
\end{equation} where \(C\) is again used to adjust for the scaling
imbalance between \(q^*\) and \(v\).

Henceforth, we assume that \(q^*\) and \(v\) are both sampled using
\(T\) points, making the dimensionality of \(g^C\in\real^{2T}\). Then,
given a sample of amplitude-phase functions \(\{g^C_1,\dots,g^C_n\}\),
and their sample mean
\(\hat{\mu}_g^C = [\hat{\mu}_{q^*}~~ \hat{\mu}_v^C]\), we can compute
the sample covariance matrix as \begin{equation}
  K_g^C = \frac{1}{n-1}\sum_{i=1}^n (g^C_i - \hat{\mu}_g^C)(g^C_i - \hat{\mu}_g^C)^\T \in \mathbb{R}^{(2T) \times (2T)}\ .
\end{equation} Taking the Singular Value Decomposition,
\(K_g^C=U_g^C\Sigma_g^C (V_g^C)^\T\), we calculate the joint principal
directions of variability in the given amplitude-phase functions using
the first \(p\leq n\) columns of \(U_g^C\). These can be converted back
to the original representation spaces (\(\mathcal{F}\) and \(\gamma\))
using the mappings defined earlier. Moreover, one can calculate the
observed principal coefficients as \(\inner{g^C_i}{U^C_{g,j}}\), for the
\(i^{th}\) function with the \(j^{th}\) principal component. The
superscript of \(C\) is used to denote the dependence of the principal
coefficients on the scaling factor.

This framework can be used to visualize the joint principal geodesic
paths. First, the matrix \(U^C_g\) is partitioned into the pair
\((U^C_{q^*},U^C_v)\). Then, the amplitude and phase paths within one
standard deviation of the mean are computed as

\begin{eqnarray}
q^{*C}_{\tau,j} &=& \hat{\mu}_{q^*} + \tau \sqrt{\Sigma^C_{g,jj}}U^C_{q^*,j} \label{eq:q} \\
v^C_{\tau,j} &=& \tau \frac{\sqrt{\Sigma^C_{g,jj}}}{C} U^C_{v,j}~,
\label{eq:v}
\end{eqnarray} \noindent where \(\tau\in \real\), \(\Sigma_{g,jj}\) and
\(U^C_{j}\) are the \(j^{th}\) principal component variance and
direction of variability, respectively (note that the mean
\(\hat{\mu}_v^C\) is always zero). Then, one can obtain a joint
amplitude-phase principal path by composing \(f^{*C}_{\tau,j}\) (this is
the function corresponding to SRSF \(q^{*C}_{\tau,j}\)) with
\(\gamma^C_{\tau,j}\) (this is the warping function corresponding to
\(v^C_{\tau,j}\)).

The results of the above procedure clearly differ for variations of
\(C\). For example, using small values of \(C\), the first few principal
directions of variability will capture more amplitude variation, while
for large values of \(C\), the leading directions reflect more phase
variation. \citet{Lee:2017} present a data-driven method for estimating
\(C\) for a given sample of functions. We use this approach in the
current work to determine an appropriate value of \(C\).

\subsection{Elastic Functional Principal Component Regression
Model}\label{elastic-functional-principal-component-regression-model}

The regression model then is \begin{equation}
    y = \alpha + \sum_{j=1}^{n_o} \inner{x_i(t)}{\xi_j(t)}b_j
    \label{eq:fPCR_elastic}
\end{equation} and can be found by solving \begin{equation}
    \{\alpha^*,\mathbf{b}^*\} = \argmin_{\alpha,\mathbf{b}} \sum_{i=1}^n |y_i - \alpha - \sum_{j=1}^{n_o} \inner{x_i(t)}{\xi_j(t)}b_j|^2,
    \label{eq:fPCR_SSE2}
\end{equation} where the appropriate function is substituted in for
\(x_i\) and appropriate eigenfunction for \(\xi_j\) from Table
\ref{tab:domains} depending on which fPCA is used for the regression.

The solution of finding the optimal \(\alpha^*\) and \(\mathbf{b}^*\) is
found using ordinary least squares. Define \(Z = [\mathbf{1}~ \Theta]\),
where \(\mathbf{1}\) is a vector of ones and \(\y = [y_1,\dots,y_n]^\T\)
and \(\Theta\in\real^{N\times n_o}\) is the matrix containing the
principal coefficients for the \(N\) samples for \(n_o\) principal
components. Then the solution for \(\alpha^*\) and \(\mathbf{b}^*\) is
\[[\alpha^*,\mathbf{b}^*]^\T = (Z^\T Z)^{-1}Z^\T\mathbf{y}.\]

\section{Elastic Functional Logistic Regression}\label{logistic}

We now develop the logistic version of the Elastic fPCR model. This
model is an extension of the linear regression model with the
appropriate link function.

\subsection{Functional Logistic Regression
Model}\label{functional-logistic-regression-model}

Let \(\{f_i\}\) denote observations of a predictor function variable and
let \(y_i\in\{-1,1\}\), for \(i=1,\dots,n\) be the corresponding binary
response variable. We define the probability of the function \(f_i\)
being in class 1 (\(y_i = 1\)) as
\[P(y_i=1|f_i) = \frac{1}{1+\exp\left(-\left[\alpha+\int_0^1 f_i(t)\beta(t)\,dt\right]\right)}.\]
This is nothing but the logistic link function
\(\phi(t) = 1/(1+\exp(-t))\) applied to the conditional mean in a linear
regression model: \(\alpha+\int_0^1 f_i(t)\beta(t)dt\)
(\citet{james:2002}). Using this relation, and the fact that
\(P(y=-1|f_i) = 1-P(y=1|f_i)\), we can express the data likelihood as:
\[ \pi( \{y_i \}| \{f_i\}, \alpha,\beta) = \prod_{i=1}^n \frac{1}{1+\exp\left(-y_i\left[\alpha+\int_0^1 f_i(t)\beta(t)\,dt\right]\right)}.\]

Assuming we observe a sequence of i.i.d. pairs
\(\{f_i(t), y_i\}, i = 1, \cdots, n\), the model is identified by
maximizing the log-likelihood according to,
\[\{\alpha^*,\beta^*\} = \argmax_{\alpha,\beta(t)} \left( \log \pi( \{y_i \}| \{f_i\}, \alpha,\beta) \right) .\]
This optimization has been the main focus of the current literature (see
e.g.,
\citet{ramsay-silverman-2005}, \citet{cardot-ferraty-sarda:2003}, \citet{Hall2007}).

\subsection{Elastic fPCR Logistic
Regression}\label{elastic-fpcr-logistic-regression}

Now consider the situation where functional predictors can include phase
variability as well as the amplitude variability. We will use the
Elastic fPCR method with the logistic link function
\begin{flalign*}
	\pi( \{y_i \}| \{f_i\}, \alpha,\mathbf{b}) =  \prod_{i=1}^n \frac{1}{1+\exp\left(-y_i\left[\alpha+\sum_{j=1}^{n_o} \inner{x_i(t)}{\xi_j(t)}b_j\right]\right)}
\end{flalign*}
where the appropriate fPCA model is used for the proper variability.

The optimization over \(\alpha\) and \(\mathbf{b}\) is found by
maximizing the log-likelihood. We can combine all the parameters --
intercept \(\alpha\) and coefficients \(b_i\)s -- in a vector form
\(\bm{\theta} = [\alpha,b_1,\dots,b_{n_o}]^\T\). Let
\(\mathbf{z}_i = [1, \inner{x_i}{\xi_1(t)},\dots,\inner{x_i}{\xi_{n_o}(t)}]^\T\).
The optimal parameter vector is given as follows: \begin{equation}
\bm{\theta}^* = \argmax_{\bm{\theta} \in \real^{p+1}}\sum_{i=1}^n \log\left(\phi\left(y_i \bm{\theta}^\T\mathbf{z}_i\right)\right),
\label{eq:logistic_ll2}
\end{equation} There is no analytical solution to this optimization
problem. Since the objective function is concave, we can use a numerical
method such as Conjugate Gradient or the
Broyden--Fletcher--Goldfarb--Shanno (BFGS) algorithm
(\citet{mordecai:2003}). To use these algorithms we need the gradient of
the log-likelihood, \(L\), which is given by:
\[\nabla L(\bm{\theta}) = \sum_{i=1}^n -y_i\mathbf{z}_i(\phi(y_i\bm{\theta}^\T\mathbf{z}_i)-1).\]
In this paper we will use the Limited Memory BFGS (L-BFGS) algorithm due
to its low-memory usage for large number of predictors
(\citet{Byrd:1995}). Similar to ideas discussed in
\cite{gerheiss-maity-staicu:2013}, one can also seek a sparse
representation by including a \(\mathbb{L}_1\) or \(\mathbb{L}_2\)
penalty on \(\mathbf{b}\) in Eqn \ref{eq:logistic_ll2}.

\subsection{Extension to Elastic fPCR Multinomial Logistic
Regression}\label{extension-to-elastic-fpcr-multinomial-logistic-regression}

We can extend the elastic functional logistic regression to the case of
multinomial response, i.e. \(y_i\) has more than two classes. In this
case, we have observations \{(\(f_i(t)\), \(y_i\))\} and the response
variable can take on \(m\) categories, \(y_i\in\{1,\dots,m\}\), for
\(i=1,\dots, n\). For simplification, we abuse the notation by coding
the response variable \(y\) as a \(m\)-dimensional vector with a 1 in
the \(k\)th component when \(y=k\) and zero, otherwise. Next, let's
define the probability of the function \(f\) being in class \(k\) as
\begin{eqnarray*}
P(y^{(k)}=1|\{\alpha^{(j)}\},\{\mathbf{b}^{(j)}\},f) = \frac{\exp\left(\alpha^{(k)}+\sum_{j=1}^{n_o} \inner{x(t)}{\xi_j(t)}b_j^{(k)}\right)}{1+\sum_{l=1}^{m-1} \exp\left(\alpha^{(l)}+\sum_{j=1}^{n_o} \inner{x(t)}{\xi_j(t)}b_j^{(l)}\right)}
\end{eqnarray*}
we only need \(m-1\) \(\alpha\)'s and \(\mathbf{b}\)'s as we can assume
\(\alpha^{(m)} = 0\) and \(\mathbf{b}^{(m)}=\mathbf{0}\) without loss of
generality.

Using the above probability and the multinomial definition of the
problem, we can express the log-likelihood of observations
\{(\(x_i(t)\), \(y_i\))\} as
\begin{eqnarray*}
L_m(\{\alpha^{(i)}\},\{\mathbf{b}^{(i)}\}) &=& \sum_{i=1}^n\left[\sum_{k=1}^{m-1} y_i^{(k)}\left[\alpha^{(k)} + \sum_{j=1}^{n_o} \inner{x_i(t)}{\xi_j(t)}b_j^{(k)}\right] \right. \\ && \left.-\log\left(1+\sum_{l=1}^{m-1} \exp\left(\alpha^{(l)}+\sum_{j=1}^{n_o} \inner{x_i(t)}{\xi_j(t)}b_j^{(l)}\right)\right)\right].
\end{eqnarray*} where again the appropriate fPCA model is used for the
proper variability modeling.

The optimal \(\{\alpha^{*(i)}\}\) and \(\{\mathbf{b}^{*(i)}\}\) will be
found again by maximizing the log-likelihood. We can re-express the
maximization problem of the log-likelihood as \begin{equation}
\bm{\theta}^* = \argmax_{\bm{\theta}}\sum_{i=1}^n \left[\sum_{j=1}^{m-1} y_i^{(j)}\bm{\theta}^{(j)\T} \mathbf{z}_i - \log\left(1+\sum_{j=1}^{m-1} \exp\left(\bm{\theta}^{(j)\T} \mathbf{z}_i\right)\right)\right],
\label{eq:mlogistic_ll2}
\end{equation} where
\(\bm{\theta} = [\alpha^{(k)},b^{(k)}_1,\dots,b^{(k)}_{n_o}]^\T\) and
\(\mathbf{z}_i = [1, \inner{x_i}{\xi_1(t)},\dots,\inner{x_i}{\xi_{n_o}(t)}]^\T\).
There is no direct solution to solving this optimization and it has to be
performed numerically. Since, the function is concave we will use the
L-BFGS algorithm to find the solution numerically. To use this algorithm
we need the gradient of the log-likelihood. We need to find the partial
derivative of the log-likelihood for each \(\mathbf{b}^{(k)}\),
\[\frac{\partial L_m(\bm{\theta})}{\partial \bm{\theta}^{(k)}} = \sum_{i=1}^n \left[y_i^{(k)} \mathbf{z}_i - \frac{1}{1+\sum_{j=1}^{m-1} \exp\left(\bm{\theta}^{(j)\T} \mathbf{z}_i\right)} \exp\left(\bm{\theta}^{(k)\T}\mathbf{z}_i\right)\mathbf{z}\right].\]
We can then find the optimal \(\{\alpha^{*(j)}\}\) and
\(\{\mathbf{b}^{*(j)}\}\) using L-BFGS.

\section{Simulation Results}\label{sim}

\subsection{Elastic Functional Principal Component
Regression}\label{elastic-functional-principal-component-regression}

\begin{figure}

{\centering \subfloat[Original Functions\label{fig:simul_data_comb1}]{\includegraphics{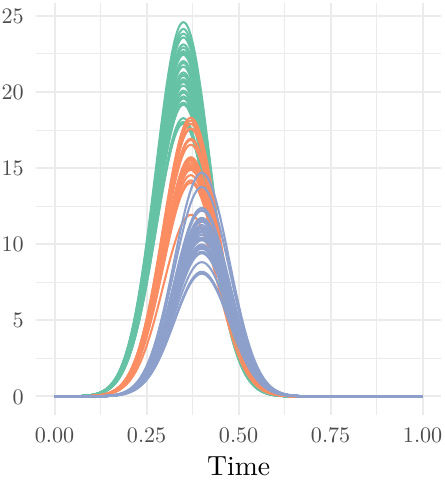} }\subfloat[Warped Functions\label{fig:simul_data_comb2}]{\includegraphics{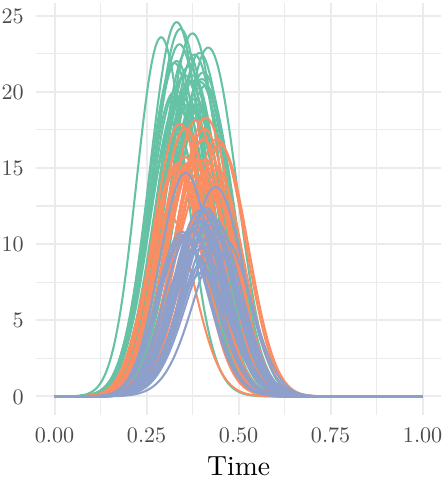} }\subfloat[Response Variable\label{fig:simul_data_comb3}]{\includegraphics{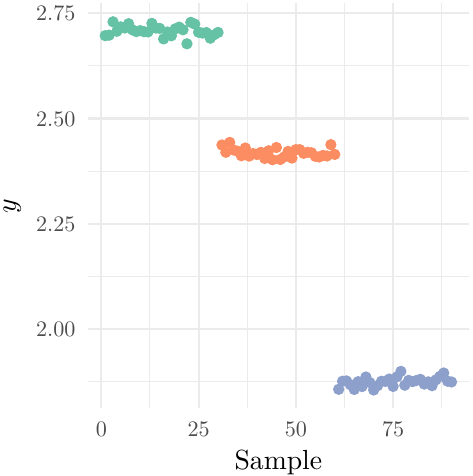} }

}

\caption{Simulated regression data with phase and amplitude variability.}\label{fig:simul_data_comb}
\end{figure}

\begin{figure}

{\centering \subfloat[Original Functions\label{fig:simul_data_vert1}]{\includegraphics{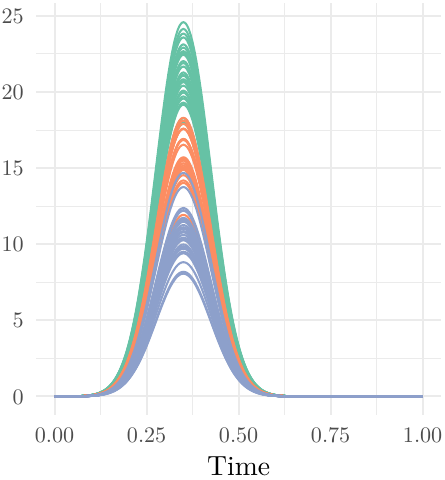} }\subfloat[Warped Functions\label{fig:simul_data_vert2}]{\includegraphics{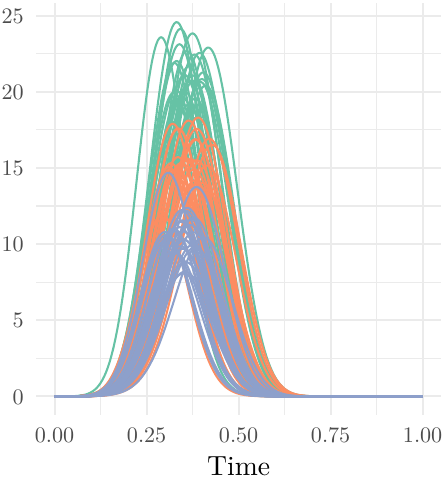} }\subfloat[Response Variable\label{fig:simul_data_vert3}]{\includegraphics{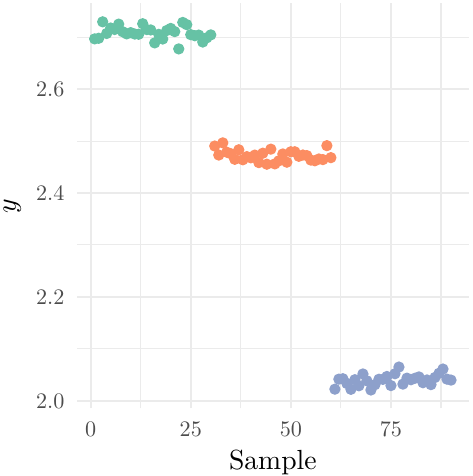} }

}

\caption{Simulated regression data with amplitude variability.}\label{fig:simul_data_vert}
\end{figure}

\begin{figure}

{\centering \subfloat[Original Functions\label{fig:simul_data_horiz1}]{\includegraphics{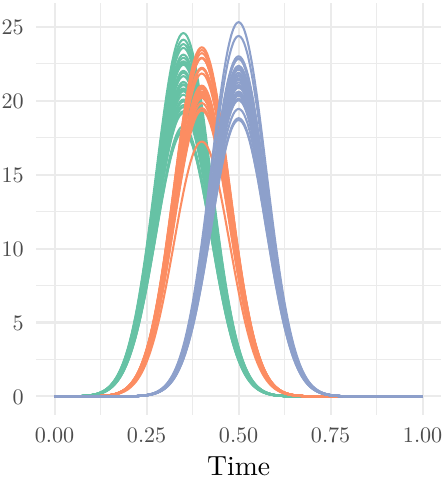} }\subfloat[Warped Functions\label{fig:simul_data_horiz2}]{\includegraphics{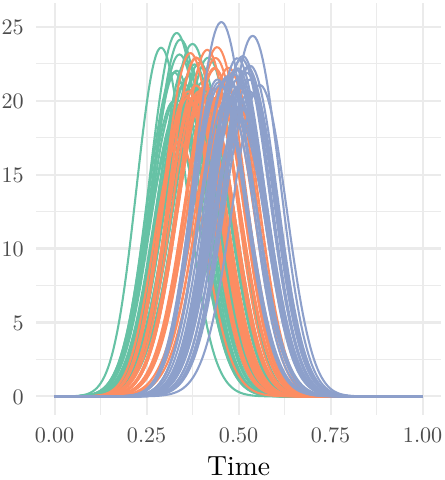} }\subfloat[Response Variable\label{fig:simul_data_horiz3}]{\includegraphics{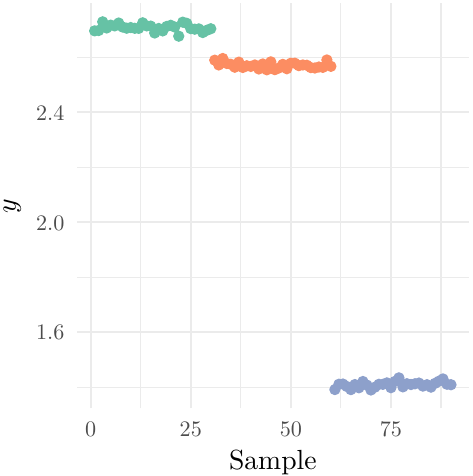} }

}

\caption{Simulated regression data with phase variability.}\label{fig:simul_data_horiz}
\end{figure}

To illustrate the developed elastic functional regression method we
evaluated the model on a simulated data constructed using
\[f_i(t) = a_i \frac{1}{\sqrt{2\pi\sigma^2}}\exp\left(-\frac{(t-\mu_j)^2}{2\sigma^2}\right),\]
where \(a_i\sim\mathcal{N}(d_j,0.05)\). The means were chosen according
to three models: 1) Combined Amplitude \& Phase Variability
(\(\mu_j\in[0.35, 0.37, 0.40]\) and \(d_j\in[4, 3, 2]\)), 2. Amplitude
Variability (\(\mu_j\in[0.35, 0.35, 0.35]\) and \(d_j\in[4, 3.7, 4]\)),
and 3) Phase Variability (\(\mu_j\in[0.35, 0.40, 0.50]\) and
\(d_j\in[4, 4, 4]\)). A total of 20 functions were generated for each
case and \(\sigma = 0.075\). The generated functions are shown in Fig.
\ref{fig:simul_data_comb}(a), \ref{fig:simul_data_vert}(a), and
\ref{fig:simul_data_horiz}(a), for cases 1, 2, and 3, respectively The
functions were the randomly warped to generate the warped data,
\(\{f_i\}\) are shown in Fig. \ref{fig:simul_data_comb}(b),
\ref{fig:simul_data_vert}(b), and \ref{fig:simul_data_horiz}(b). The
response variable \(y_i\) was generated with \(\alpha = 0\),
\(\beta(t) = 0.5\sin(2\pi t) + 0.9 \cos(2\pi t)\) and is shown in Fig.
\ref{fig:simul_data_comb}(c), \ref{fig:simul_data_vert}(c), and
\ref{fig:simul_data_horiz}(c).

\begin{table*}[!h]

\caption{\label{tab:simul_pcr_horiz}Calculated SSE values using four different functional PCR methods for 3 different types of variability.}
\centering
\begin{tabular}[t]{lcccc}
\toprule
  & Elastic Combined & Elastic Vertical & Elastic Horizontal & Standard\\
\midrule
Combined & 0.0875 (0.0237) & 0.3498 (0.1248) & \textbf{0.0838 (0.0300)} & 0.5075 (0.1483)\\
Vertical & \textbf{0.2345 (0.0905)} & 0.2390 (0.0869) & 2.0782 (1.9058) & 0.3173 (0.0393)\\
Horizontal & \textbf{0.1474 (0.0998)} & 9.1292 (5.3023) & 0.2155 (0.1339) & 1.8729 (1.1719)\\
\bottomrule
\end{tabular}
\end{table*}

Table \ref{tab:simul_pcr_horiz}provides the SSE for each of the three
cases with the lowest SSE shown in bold for the applied fPCR method. For
the data with the combined variability the combined fPCA in the elastic
fPCR model is slightly out performed by the horizontal fPCA. In the cases
with the vertical and horizontal variability, the combined elastic fPCA
method performed the best with the corresponding vertical or horizontal
fPCA method being very close. This is somewhat to be expected as the
combined fPCA method is able to capture both types of variability. We
compared the results from the elastic method to those using standard
fPCR found in the literature on the warped data and is shown in the last
column. In all cases the elastic method outperforms the standard fPCR
method presented by \citet{reiss-ogden:2007}.

\subsection{Elastic Logistic fPCR}\label{elastic-logistic-fpcr}

To illustrate the developed elastic functional logistic regression
method, we evaluated the model on a similar simulated data used in the
previous section. The means were chosen according to three models: 1)
Combined Amplitude \& Phase Variability (\(\mu_j\in[0.35, 0.37]\) and
\(d_j\in[4, 3]\)), 2. Amplitude Variability (\(\mu_j\in[0.35, 0.35]\)
and \(d_j\in[4, 3.7]\)), and 3) Phase Variability
(\(\mu_j\in[0.35, 0.40]\) and \(d_j\in[4, 4]\)). A total of 20 functions
were generated for each case and \(\sigma = 0.075\). The functions were
then randomly warped to generate the warped data; \(\{f_i\}\) and the
label was 1 for the first case and -1 for the second case.

Table \ref{tab:simul_lpcr}provides the combined probability of
classification (PC) for each of the three cases. For the data with the
combined variability the combined fPCA in the elastic logistic fPCR
model performed the best. In the cases with the vertical and horizontal
variability, the corresponding elastic fPCA methods performed well with
the combined fPCA method and showed the best performance. We compared the
results to the standard logistic fPCR, where the logistic link function
is applied to the method of \citet{reiss-ogden:2007}. The results are
shown in the last column. In all cases the elastic method outperforms
the standard logistic fPCR method.

\begin{table*}[!h]

\caption{\label{tab:simul_lpcr}Calculated combined probability of classification values using four different functional PCR methods, for 3 different types of variability for logistic regression.}
\centering
\begin{tabular}[t]{lcccc}
\toprule
  & Elastic Combined & Elastic Vertical & Elastic Horizontal & Standard\\
\midrule
Combined & \textbf{0.9750 (0.0342)} & 0.9375 (0.0765) & 0.9750 (0.0342) & 0.9625 (0.0342)\\
Vertical & \textbf{0.9250 (0.0280)} & 0.8500 (0.0948) & 0.6250 (0.0625) & 0.8750 (0.0442)\\
Horizontal & \textbf{0.9250 (0.0815)} & 0.6000 (0.1630) & 0.8875 (0.1355) & 0.8750 (0.0442)\\
\bottomrule
\end{tabular}
\end{table*}

\subsection{Elastic Multinomial Logistic
fPCR}\label{elastic-multinomial-logistic-fpcr}

To illustrate the developed elastic functional regression method we
evaluated the model on the simulated data constructed in the elastic
functional PCR case. Each of the functions was randomly warped similar
to the previous cases. The response variable \(y_i\) in this case was
categorical with values \(j\in\{1,2,3\}\) depending on the corresponding
model.

Table \ref{tab:simul_mlpcr}provides the combined probability of
classification (PC) for each of the three cases. For the data with the
combined variability the horizontal and combined fPCA in the elastic
multinomial logistic fPCR model performed the best. In the cases with the
vertical and horizontal variability, the corresponding elastic fPCA
methods performed the well with the combined fPCA method having the best
performance. We compared the results to using standard multinomial
logistic fPCR found in the literature on the warped data and is presented in
the last column. In all cases the elastic method outperforms the
standard multinomial logistic fPCR method.

\begin{table*}[!h]

\caption{\label{tab:simul_mlpcr}Calculated combined probability of classification values using four different functional PCR methods, for 3 different types of variability for multinomial logistic regression.}
\centering
\begin{tabular}[t]{lcccc}
\toprule
  & Elastic Combined & Elastic Vertical & Elastic Horizontal & Standard\\
\midrule
Combined & 0.9420 (0.0633) & 0.9246 (0.0355) & \textbf{0.9670 (0.0348)} & 0.8663 (0.1597)\\
Vertical & 0.8589 (0.0801) & \textbf{0.8916 (0.0229)} & 0.3822 (0.0710) & 0.8749 (0.0780)\\
Horizontal & 0.9176 (0.0480) & 0.3822 (0.1036) & \textbf{0.9666 (0.0187)} & 0.9510 (0.0436)\\
\bottomrule
\end{tabular}
\end{table*}

\section{Applications to Real Data}\label{realdata}

Here, we present the results on multiple real data sets for the three
elastic regression models. For the elastic fPCR we use the Sonar data
set presented in (\citet{tucker-wu-srivastava:2013b}) where we predict
the volumes of two targets. We demonstrate the elastic logistic fPCR
model on four sets. The data consists of physiological data,
specifically, gait and electrocardiogram (ECG) measurements from various
patients. Phase-variability is naturally found in the data, as during
collection the signals always start and stop at the different time for
each measurement. For example, when measuring a heart beat one cannot
assure that the measurement starts on the same part of the heartbeat for
each patient measured. For the elastic multinomial logistic fPCR model
we demonstrate on two sets that consist of physiologic data similar to
those used to test the logistic regression method.

\subsection{Sonar Data}\label{sonar-data}

The data set used in these experiments was collected at the Naval
Surface Warfare Center Panama City Division (NSWC PCD) test pond. For a
description of the pond and a similar experimental setup the reader is
referred to \citet{art:kargl}. The raw SONAR data was collected using a
1 - 30\(kHz\) LFM chirp and data was collected for a solid aluminum
cylinder and an aluminum pipe. The aluminum cylinder is 2\(ft\) long with
a 1\(ft\) diameter; while the pipe is 2\(ft\) long with an inner
diameter of 1\(ft\) and 3/8 inch wall thickness. During the experiment
the targets were placed with added uncertainty of their orientation. The
acoustic signals were generated from the raw SONAR data to construct
target strength as a function of frequency and aspect angle.

\begin{figure*}

{\centering \includegraphics{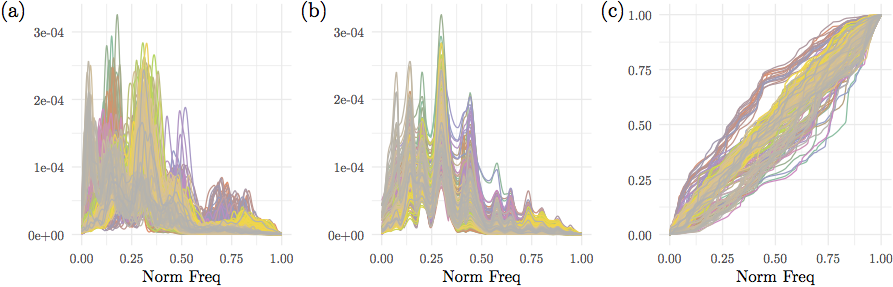}

}

\caption{Alignment of the sonar dataset. (a) Original functions. (b) Aligned functions (amplitude). (c) Warping functions (phase).}\label{fig:sonar_data}
\end{figure*}

Figure \ref{fig:sonar_data}(a) presents the original functions for the
acoustic color measurements at \(0^\circ\) aspect angle. There appears
to be significant amplitude and phase variability between functional
measurements due to experimental collection uncertainty. Not accounting
for the phase variability can greatly affect summary statistics and
follow-on statistical models. Figure \ref{fig:sonar_data}(b) and (c)
show the aligned functions (amplitude) and warping functions (phase),
respectively. Overall there is significant difference between the
original functions and the aligned functions. With the large amount of
phase variability, the frequency structure of the data was lost. As a
result, cross-sectional methods without alignment will not capture this
important difference in the functions.

\begin{table*}[!h]

\caption{\label{tab:sonar_pcr}Calculated SSE values using four different functional PCR methods for the sonar data set.}
\centering
\begin{tabular}[t]{lcccc}
\toprule
  & Elastic Combined & Elastic Vertical & Elastic Horizontal & Standard\\
\midrule
SSE & \textbf{0.1210 (0.0472)} & 0.1497 (0.1399) & 0.1597 (0.0485) & 0.1908 (0.1184)\\
\bottomrule
\end{tabular}
\end{table*}

Table \ref{tab:sonar_pcr}presents the sum of squared errors (SSE)
calculated using 5-fold cross-validation. For this data set, we use ten
principal components resulting in a ten-dimensional model for all four
methods. In the table we present the mean of the SSE across the folds,
along with the standard deviation. We compare the three elastic versions
and standard functional principal component regression; the lowest
SSE is shown in bold. The lowest SSE is the combined elastic fPCR method
and all three elastic methods have lower SSE than the standard method in
predicting the volume from the sonar data. With the high degree of phase
and amplitude variability in the data, the elastic method is more able to
accurately predict and capture the variability.

\subsection{Gait Data}\label{gait-data}

The Gait data is a collection of gait measurements for patients having
Parkinson's disease, and those not having Parkinson's disease. It is from
the gaitpdb data set on Physionet (\citet{physiobank}). This database
contains measures of gait from 93 patients with idiopathic Parkinson's
disease and 73 healthy patients. The gait was measured using vertical
ground reaction force records of subjects as they walked at their usual,
self-selected pace for approximately 2 minutes on level ground.

\begin{figure*}

{\centering \includegraphics{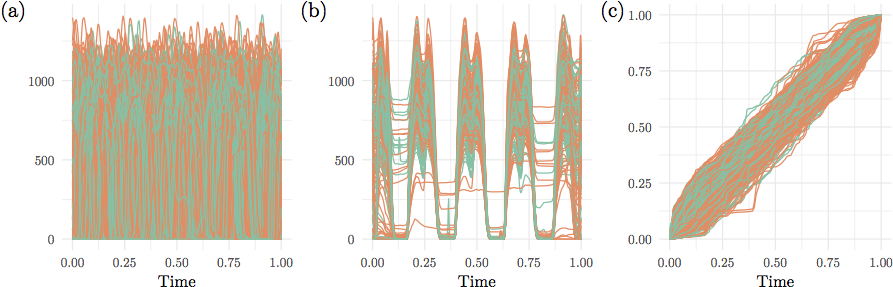}

}

\caption{Alignment of the gait data dataset. (a) Original functions. (b) Aligned functions (amplitude). (c) Warping functions (phase).}\label{fig:gait_data}
\end{figure*}

Figure \ref{fig:gait_data}(a) presents the original functions for gait
data and are colored for the two different classes. There appears to be
significant amplitude and phase variability between functional
measurements due to experimental collection uncertainty and where one
subject will start and stop their gate. Figure \ref{fig:gait_data}(b)
and (c) show the aligned functions (amplitude) and warping functions
(phase), respectively. Overall there is significant difference between
the original functions and the aligned functions. With the large amount
of phase variability, the temporal structure of the gaits will be lost
in the analysis. As a result, cross-sectional methods without alignment
will not capture this important difference in the functions.

The first row in Table \ref{tab:elastic_lpcr_data}presents the
calculated mean probability of classification (PC) using 5-fold
cross-validation. For this data set, we use five principal components
resulting in a five-dimensional model for all four methods. In the table
we present the mean of the PC across the folds, along with the standard
deviation. We compare the three elastic versions and standard logistic
fPCR and the largest PC is shown in bold. The largest PC is the vertical
elastic logistic fPCR method. All three elastic methods have higher
PC than the standard method for predicting if the subject has Parkinson's
based on gait measurement. This suggests that a large portion of the
information is contained in the amplitude variability.

\subsection{ECG200 Data}\label{ecg200-data}

The ECG200 data is a collection of ECG measurements of heartbeats
demonstrating an arrhythmia, and those which do not. The data set
is from the MIT-BIH Arrhythmia Database available from Physionet. The
database contains ECG recordings where each electrocardiogram was
recorded from a single patient for a duration of approximately thirty
minutes. From the recordings heartbeats were extracted with the most
prevalent abnormality---supra-ventricular premature beat. Additionally,
heartbeats were extracted from the recordings that were representative
of normal heartbeats. The task is then to distinguish between the
abnormalities using the heartbeat. Naturally, the heartbeats are not
aligned and no alignment was made to the data.

\begin{figure*}

{\centering \includegraphics{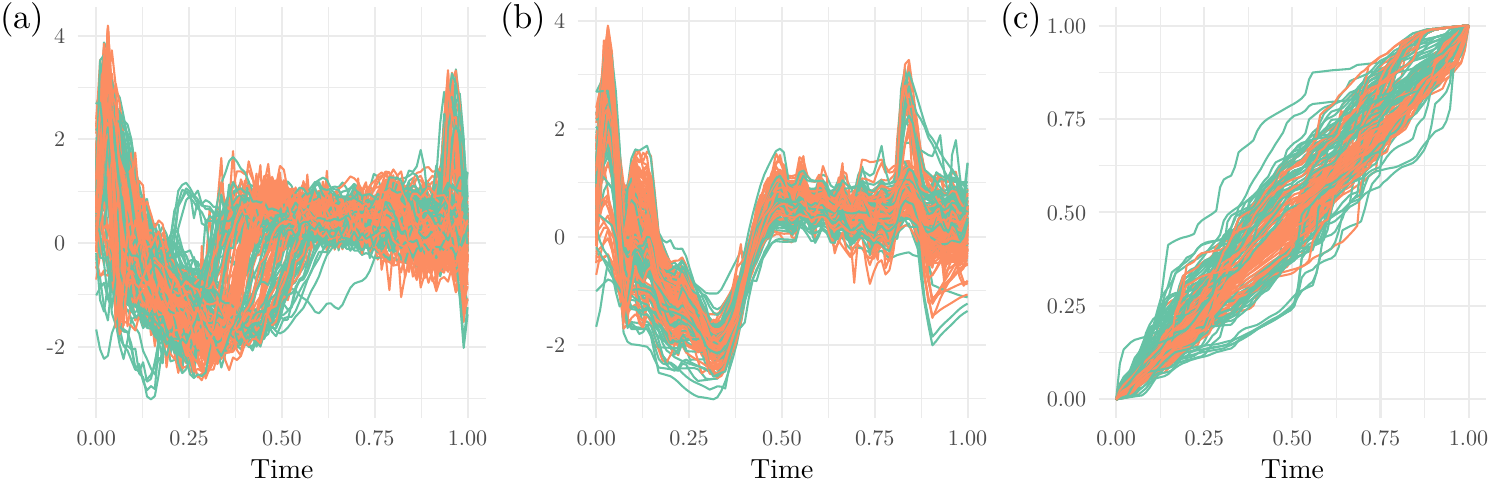}

}

\caption{Alignment of the ECG200 data dataset. (a) Original functions. (b) Aligned functions (amplitude). (c) Warping functions (phase).}\label{fig:ecg200_data}
\end{figure*}

Figure \ref{fig:ecg200_data}(a) presents the original electrocardiogram
measurements. There appears to be significant phase variability between
functional measurements due to timing uncertainty across collections.
Figure \ref{fig:ecg200_data}(b) and (c) show the aligned functions
(amplitude) and warping functions (phase), respectively.

The second row in Table \ref{tab:elastic_lpcr_data}presents the
calculated mean probability of classification (PC) using 5-fold
cross-validation. Again for this data set, we use five principal
components resulting in a five-dimensional model for all four methods,
and present the mean of the PC across the folds; along with the standard
deviation. The largest PC is the vertical elastic logistic fPCR method.
All three elastic methods have higher PC than the standard method,
however, for this data it performs quite well.

\subsection{TwoLead ECG Data Set}\label{twolead-ecg-data-set}

The TwoLead ECG data set, is a collection of ECG measurements from the
MIT-BIH Long-Term ECG Database available as well from Physionet. These
contains long term ECG measurements with beat annotations. Heartbeats
were extracted that were annotated normal and abnormal for the two
classes.

\begin{figure*}

{\centering \includegraphics{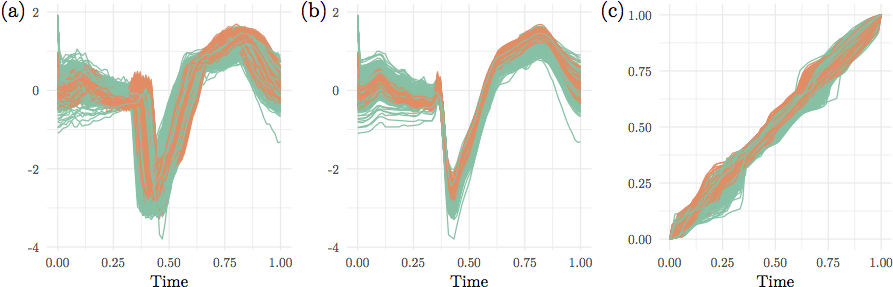}

}

\caption{Alignment of the TwoLead ECG data dataset. (a) Original functions. (b) Aligned functions (amplitude). (c) Warping functions (phase).}\label{fig:ecgtwolead_data}
\end{figure*}

Figure \ref{fig:ecgtwolead_data}(a) presents the original
electrocardiogram measurements. Again, there appears to be significant
phase variability between functional measurements due to timing
uncertainty across collections. Figure \ref{fig:ecgtwolead_data}(b) and
(c) show the aligned functions (amplitude) and warping functions
(phase), respectively. Overall there is a noticeable alignment and
better definition of the wave structure.

The third row in Table \ref{tab:elastic_lpcr_data}presents the
calculated mean probability of classification (PC) using 5-fold
cross-validation. Again for this data set, we use five principal
components resulting in a five-dimensional model for all four methods
and present the mean of the PC across the folds, along with the standard
deviation. The largest PC is the vertical elastic logistic fPCR method
and all three elastic methods have higher PC than the standard method.

\subsection{ECGFiveDays Data Set}\label{ecgfivedays-data-set}

The ECGFiveDays data set, is a collection of ECG measurements from a 67
year old male. There are two classes which are simply the data of the
ECG measurements which are 5 days apart. The task is then to distinguish
between the two days, as the wandering baseline was not removed and the
heartbeats are not aligned. The data set is the ECGFiveDays from the UCR
Time Series Classification Database (\citet{UCR}). Moreover, the
previous two data sets can also be obtained from the UCR database under
the names ECG200 and TwoLeadECG, respectively.

\begin{figure*}

{\centering \includegraphics{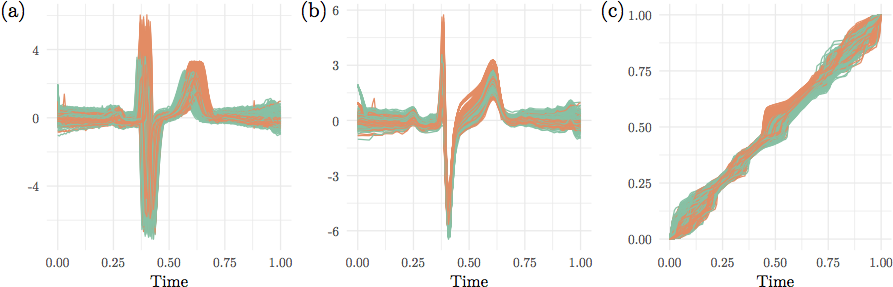}

}

\caption{Alignment of the ECGFiveDays data dataset. (a) Original functions. (b) Aligned functions (amplitude). (c) Warping functions (phase).}\label{fig:ECGFiveDays_data}
\end{figure*}

Figure \ref{fig:ECGFiveDays_data}(a) presents the original
electrocardiogram measurements from the ECGFiveDays set. Again, there
appears to be some phase variability between functional measurements due
to timing uncertainty across collections. Figure
\ref{fig:ECGFiveDays_data}(b) and (c) show the aligned functions
(amplitude) and warping functions (phase), respectively. Overall there
is a noticeable alignment and separation of the two classes in both the
aligned functions and the warping functions.

\begin{table*}[!h]

\caption{\label{tab:elastic_lpcr_data}Calculated probability of correct classification using four different functional logistic fPCR methods for four different data sets.}
\centering
\begin{tabular}[t]{lcccc}
\toprule
  & Elastic Combined & Elastic Vertical & Elastic Horizontal & Standard\\
\midrule
Gait & 0.6467 (0.0321) & \textbf{0.6900 (0.0465)} & 0.6333 (0.0425) & 0.4300 (0.0923)\\
ECG200 & 0.7750 (0.0791) & \textbf{0.8350 (0.0675)} & 0.7450 (0.0716) & 0.8200 (0.0694)\\
TwoLead ECG & 0.9113 (0.0125) & \textbf{0.9845 (0.0163)} & 0.9156 (0.0146) & 0.8012 (0.0133)\\
ECGFiveDays & \textbf{0.9570 (0.0228)} & 0.8902 (0.0462) & 0.8473 (0.0429) & 0.9061 (0.0297)\\
\bottomrule
\end{tabular}
\end{table*}

The last row in Table \ref{tab:elastic_lpcr_data}presents the
calculated mean probability of classification (PC) using 5-fold
cross-validation. Again for this data set, we use five principal
components resulting in a five-dimensional model for all four methods.
We present the mean of the PC across the folds, along with the standard
deviation. The largest PC is the combined elastic fPCR method and all
three elastic methods have higher PC than the standard method. This
suggests that there is a combination of both phase and amplitude that
contribute to correct classification.

\subsection{Gaitndd Data set}\label{gaitndd-data-set}

The Gaitndd data set is a collection of gait measurements for patients
having Parkinson's disease, Amyotrophic lateral sclerosis, Huntington's
disease, and healthy controls; and is from the gaitndd data set on
Physionet (\citet{physiobank}). This database contains measures of gait
from 15, 20, 13, and 16 patients for the respective diseases. The gait
was measured using vertical ground reaction force records of subjects as
they walked at their usual pace.

\begin{figure*}

{\centering \includegraphics{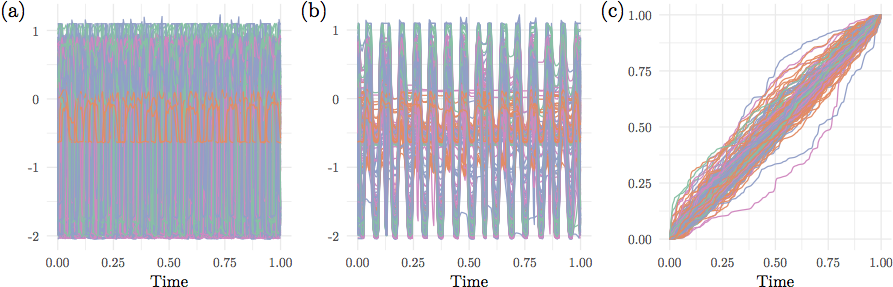}

}

\caption{Alignment of the Gaitndd data dataset. (a) Original functions. (b) Aligned functions (amplitude). (c) Warping functions (phase).}\label{fig:Gaitndd_data}
\end{figure*}

Figure \ref{fig:Gaitndd_data}(a) presents the original gait measurements
and are colored for the different classes. There is a large phase and
amplitude variability between functional measurements. Figure
\ref{fig:Gaitndd_data}(b) and (c) show the aligned functions (amplitude)
and warping functions (phase), respectively. Overall there is a large
improvement in the structure after alignment, and some class definition can
be noticed in the functions.

The first row in Table \ref{tab:elastic_mlpcr_table}presents the
calculated mean probability of classification (PC) using 5-fold
cross-validation. For this data set, we use ten principal components
resulting in a ten-dimensional model for all four methods. We then present
the mean of the PC across the folds, along with the standard deviation.
The largest PC is the vertical elastic fPCR method and all three elastic
methods have higher PC than the standard method. This suggests there is
a large amplitude component in how each disease affects the gait.

\subsection{CinC ECG Data Set}\label{cinc-ecg-data-set}

The last data set is a collection of ECG measurements from multiple
torso-surface sites. There are measurements from 4 different people
who are the 4 different classes. The data set is from the 2007
Physionet CinC challenge and is also found as the CinC data set from the
UCR Time Series Classification Database (\citet{UCR}).

\begin{figure*}

{\centering \includegraphics{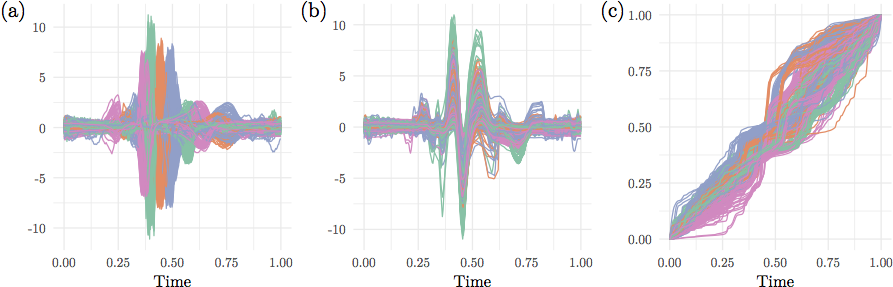}

}

\caption{Alignment of the CinC ECG data dataset. (a) Original functions. (b) Aligned functions (amplitude). (c) Warping functions (phase).}\label{fig:ecgcinc_data}
\end{figure*}

Figure \ref{fig:ecgcinc_data}(a) presents the original ECG measurements
and are colored for the different classes. There is a large phase and
amplitude variability between functional measurements. Figure
\ref{fig:ecgcinc_data}(b) and (c) show the aligned functions (amplitude)
and warping functions (phase), respectively. Overall there is a large
improvement in the structure after alignment and noticeable class
separation in the warping functions. This suggests that the phase will
have a large contribution to the classification.

\begin{table*}[!h]

\caption{\label{tab:elastic_mlpcr_table}Calculated probability of correct classification using four different functional multinomial logistic fPCR methods for two different data sets.}
\centering
\begin{tabular}[t]{lcccc}
\toprule
  & Elastic Combined & Elastic Vertical & Elastic Horizontal & Standard\\
\midrule
Gaitndd & 0.3949 (0.0648) & \textbf{0.4888 (0.0326)} & 0.3645 (0.0398) & 0.3123 (0.0413)\\
CinC ECG & 0.6785 (0.0403) & 0.6342 (0.0311) & \textbf{0.6954 (0.0452)} & 0.3297 (0.0374)\\
\bottomrule
\end{tabular}
\end{table*}

The last row in Table \ref{tab:elastic_mlpcr_table}presents the
calculated mean probability of classification (PC) using 5-fold
cross-validation. For this data set, we use ten principal components
resulting in a ten-dimensional model for all four methods and present
the mean of the PC across the folds, along with the standard deviation.
The largest PC is the horizontal elastic fPCR method, and all three
elastic methods have higher PC than the standard method. This suggests
there is a large phase component in the classification performance. When
accounting for this performance of correct classification is
dramatically larger than just performing standard multinomial functional
principal component regression.

\section{Conclusion and Future Work}\label{conclusion}

The statistical modeling and classification of functional data with
phase variability is a challenging task. We have proposed a new
functional principal component regression approach, that addresses the
problem of registering and modeling functions in one elastic-framework.
We demonstrated three PCA methods: 1) combined, 2) vertical, and 3)
horizontal that can be used depending on the type of data encountered.
This enabled the implementation of a regression model that is
geometrically-motivated. We demonstrated the applicability of these to
models on a three different simulated examples, that contain different
types of variability. We also tested seven real data examples with
significant amplitude and phase variabilities. In all cases, we
illustrated improvements in prediction power of the proposed models.

We have identified several directions for future work. First, we will
explore the influence of the weight \(C\) in the combined amplitude and
phase fPCA model on the resulting regression model performance. Second,
in many applications, the functional data of interest may be more
complex than the simple univariate functions considered in this work;
some examples include shapes of curves, surfaces, and images. These more
complicated data objects often exhibit different sources of variability,
which must be taken into account when computing regression models.

\bibliography{JDTBib.bib}
\bibliographystyle{chicago}

\end{document}